\documentclass[aps,twocolumn,prl,showpacs,preprintnumbers,amsmath,amssymb,floatfix]{revtex4}
\usepackage{graphicx}
\usepackage{dcolumn}
\usepackage{bm}

\begin{document}

\preprint{APS/123-QED}

\author{P. Gegenwart$^{(1)}$, F. Weickert$^{(1)}$, M. Garst$^{(2)}$, R.S. Perry$^{(3,4,5)}$, Y. Maeno$^{(4,5)}$}
\address{
$^{(1)}$ Max-Planck Institute for Chemical Physics of Solids,
D-01187 Dresden, Germany
\\ $^{(2)}$Theoretical Physics Institute, University of Minnesota,
Minneapolis, Minnesota 55455, USA
\\ $^{(3)}$School of Physics and Astronomy, University of St. Andrews, Fife KY16 9SS, Scotland
\\ $^{(4)}$International Innovation Center, Kyoto University, Japan
\\ $^{(5)}$Department of Physics, Kyoto University, Japan}

\title{Metamagnetic quantum criticality in Sr$_3$Ru$_2$O$_7$ studied by thermal expansion}

\date{\today}

\begin{abstract}
We report low-temperature thermal expansion measurements on the
bilayer ruthenate Sr$_3$Ru$_2$O$_7$ as a function of magnetic
field applied perpendicular to the Ruthenium-oxide planes. The
field-dependence of the $c$-axis expansion coefficient indicates
the accumulation of entropy close to 8 Tesla, related to an
underlying quantum critical point. The latter is masked by two
first-order metamagnetic transitions which bound a regime of
enhanced entropy. Outside this region the singular thermal
expansion behavior is compatible with the predictions of the
itinerant theory for a two-dimensional metamagnetic quantum
critical end point.
\end{abstract}


\maketitle

Quantum phase transitions (QPTs) in itinerant electron systems are
of extensive current interest in condensed matter physics because
they are not only at the origin of unusual finite temperature
properties but also promote the formation of new states of matter
like unconventional superconductivity in heavy fermion systems
\cite{Mathur}. Dilatometric studies are especially suitable to
investigate quantum criticality since they directly probe the
sensitivity of thermodynamics resulting from the QPT being
susceptible to pressure-tuning. In particular, near any
pressure-tuned quantum critical point (QCP) the thermal
expansion, which at constant pressure quantifies the temperature
dependence of the sample volume, is more singular than the
specific heat \cite{Zhu,Kuechler03}.  Below, we use thermal
expansion to study a certain type of QCP associated with
metamagnetism in metals. Generally, metamagnetism describes the
sudden rise of the magnetization $M(H)$ as a function of the
applied magnetic field $H$. A line of first-order metamagnetic
transitions in the magnetic field--temperature plane has an end
point ($H^*,T^*$) beyond which the transition becomes a
continuous cross-over. An especially interesting situation now
arises when the temperature $T^*$ can be tuned to zero giving
rise to an isolated quantum critical end point (QCEP)
\cite{Millis02}.

The material that initiated the interest in QCEPs is the ruthenate
Sr$_3$Ru$_2$O$_7$ \cite{Grigera01}, which crystallizes in the
bilayered version of the perovskite structure. Its metamagnetic
transition conforms to the picture of a field-induced Stoner
transition given the absence of a dramatic change of the Fermi
surface \cite{Borzi04} and its highly enhanced Sommerfeld-Wilson
ratio. Sr$_3$Ru$_2$O$_7$ shows an almost isotropic zero-field
susceptibility at low temperatures, and metamagnetism occurs both
between 5 and 6~T for fields in the Ruthenium-oxide planes as
well as near 7.8~T for $H\parallel c$ \cite{Grigera01}. A detailed
study on single crystals with a residual resistivity of about
3~$\mu\Omega$cm has revealed that the temperature, $T^*$, of the
critical end point of a line of first-order metamagnetic
transitions can indeed be tuned by changing the angle of the
applied magnetic field from 1.25~K for fields perpendicular to
the $c$-axis towards zero for $H\parallel c$ \cite{Grigera03}.
For this field orientation, non-Fermi liquid behavior occurs
related to a QCEP close to 8~T \cite{Grigera01}.  However, the
electrical resistivity at fields very close to the metamagnetic
region has revealed additional features:
In a new generation of high-quality Sr$_3$Ru$_2$O$_7$ single
crystals with a residual resistivity as low as 0.4~$\mu\Omega$cm
\cite{Perry04} three peaks in the susceptibility and
magnetostriction have been observed below 1~K. The first one at
7.5~T marks a cross-over, whereas the two latter ones at 7.85 and
8.07~T are first-order metamagnetic transitions. They confine a
low-temperature ($T\leq 1$~K) regime in which the residual
electrical resistivity is strongly enhanced indicative of strong
elastic scattering and which disappears by rotating the field for
more than 10 degrees away from $H\parallel c$ \cite{Grigera04}. It
has been proposed \cite{Grigera04} that this behavior might
result from the formation of a symmetry-broken phase
characterized by a spin-dependent Pomeranchuk deformation of the
Fermi surface. Alternatively, a scenario of phase separation
leading to charge inhomogeneities within the bounded regime has
been discussed \cite{Honerkamp05}.

In this Letter, we report a $c$-axis thermal expansion study on
Sr$_3$Ru$_2$O$_7$ at temperatures down to 100~mK and mT field
steps close to the metamagnetic regime for $H\parallel c$. We
observe quantum critical behavior compatible with the predictions
for a two-dimensional (2D) metamagnetic QCEP \cite{Millis02}.
Distinct anomalies in thermal expansion and electrical
resistivity at the bounded regime are discussed.

The thermal expansion measurements were performed on a
high-quality single crystal with 1.7~mm length along the
$c$-axis, grown by floating zone technique \cite{Perry04}. Length
changes along the $c$-axis have been detected by utilizing a
high-resolution capacitive dilatometer in a dilution
refrigerator. Below, we use nominal values for the field applied
with aid of a 20~T superconducting magnet. For comparison with
corresponding susceptibility measurements \cite{Grigera04} they
need to be offset by 0.037~T. The linear thermal expansion
coefficient $\alpha_c=d(\Delta L_c/L_c)/dT$ is obtained by
calculating the slope of the relative $c$-axis length change in
temperature intervals of 40~mK. Additionally, the electrical
resistivity of a small plate-like piece from the same batch has
been measured using a low-frequency four-point ac-method. The
residual resistivity ratio of this sample equals 150.

\begin{figure}[t]
\includegraphics[width=\linewidth,keepaspectratio]{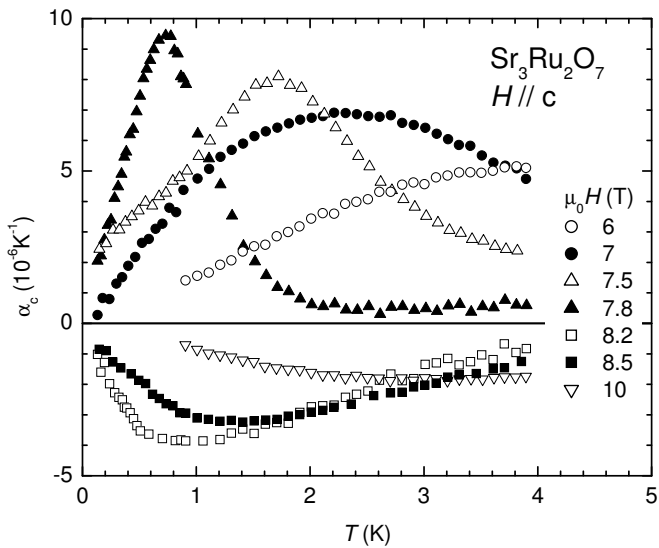}
\caption{\label{fig1}Linear $c$-axis thermal expansion
coefficient of Sr$_3$Ru$_2$O$_7$ as $\alpha_c$ vs $T$ at various
magnetic fields $H\parallel c$ indicated by different symbols.}
\end{figure}

Fig.~\ref{fig1} shows the temperature dependence of $\alpha_c$ in
different fields between 6 and 10~T for temperatures up to 4~K.
Strikingly, the thermal expansion changes sign across the
metamagnetic cross-over. This was also observed near the
metamagnetic cross-over in the heavy-fermion compound
CeRu$_2$Si$_2$ \cite{Lacerda89}. Moreover, as the critical field
$H_c$ is approached the pronounced peak in $\alpha_c$ increases,
narrows and shifts towards lower temperatures.

\begin{figure}[t]
\includegraphics[width=\linewidth,keepaspectratio]{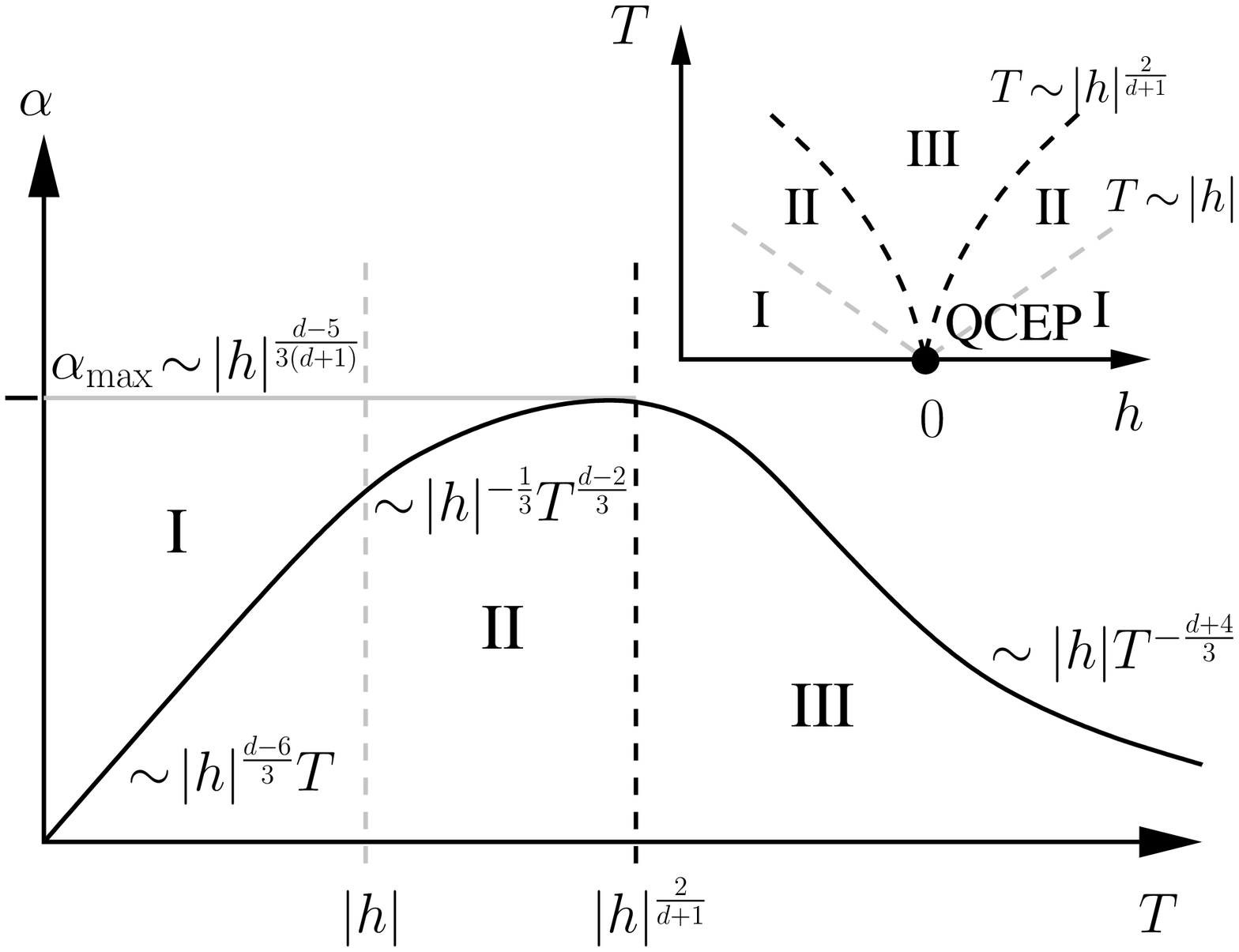}
\caption{\label{fig:Theory} Behaviour of the thermal expansion,
$\alpha$, in $d$ dimensions near the metamagnetic QCEP of
(\ref{action}), as a function of temperature, $T$, and distance
to the critical field, $h \propto H-H_c$. As explained in the
text, $\alpha(-h) = -\alpha(h)$.  Note that the intermediate
regime II shrinks to zero with $h$. In $d=2$ additional
logarithmic corrections are present; II: $\alpha \sim |h|^{-1/3}
\log \frac{T}{|h|}$, III: $\alpha \sim
h/(T^{2}\log\frac{1}{T})$.  }
\end{figure}

The characteristic shape of the thermal expansion as a function
of $T$ and $H$ can be understood within the scenario of a QCEP. As
outlined in Ref.~\cite{Millis02}, near the critical magnetic
field, $H_c$, thermodynamics is governed by the longitudinal
fluctuations of the magnetic polarization of the Fermi sea
described by an Ising field $\psi$ with the action
\begin{align} \label{action}
\mathcal{S}[\psi] &= \int \frac{d^d k}{(2\pi)^d}
\frac{1}{\beta}\sum_{\omega_n}
 \frac{1}{2} \left[r_0 + k^2 + \frac{|\omega_n|}{k}
\right] |\psi(\omega_n,k)|^2
\nonumber\\
&+ \int_0^\beta d\tau \int d^dx \left[\frac{u}{4!} \psi^4(\tau,x)
- h \psi(\tau,x)\right] \,.
\end{align}
When the system is tuned to its QCEP, $r_0\approx 0$, which will
be assumed in the following, these fluctuations are very soft
near the metamagnetic transition resulting in strong
thermodynamic signatures. The deviation of the magnetic field
from its critical value, $h \propto H - H_c$, then acts as the
control parameter of metamagnetism.

The leading contributions to the thermal expansion, $\alpha =
V_m^{-1} (dV/dT)_{p,H} = - V_m^{-1} (dS/dp)_{T,H}$ ($V_m$: molar
volume) will generically derive from the pressure dependence of
the most relevant coupling, i.e.~$h$ or, equivalently, the
critical magnetic field, $H_c = H_c(p)$. Derivatives of the free
energy with respect to $H$ and $p$ then probe the same
thermodynamic information, which would naturally explain the
close correspondence between magnetostriction and susceptibility
observed in Sr$_3$Ru$_2$O$_7$ \cite{Grigera04}.  For the critical
part of the thermal expansion we get the relation $\alpha =
\Omega V_m^{-1} (\partial S/\partial H)_{T}$, where $\Omega
\equiv (dH_c/dp)_{H=H_c}\approx 5.6$~T/Gpa \cite{Chiao} is a
measure for the (linear) hydrostatic pressure dependence of $H_c$.
The corresponding relation also holds for the $c$-axis thermal
expansion and the $c$-axis uniaxial pressure dependence of the
critical field, which will explain the observed sign change of
$\alpha_c$ across the metamagnetic transition, see
Fig.~\ref{fig1}. Indeed, as detailed in \cite{Garst-Rosch} a
change of sign of the thermal expansion is a generic phenomenon
near QCPs. The soft fluctuations associated with the $T=0$
critical point enhance the entropy at finite $T$; the entropy
will therefore increase when $H_c$ is approached, which is
reflected in a sign change of $\alpha \propto (\partial
S/\partial H)_{T}$ across the transition. In particular, the
accumulation point of entropy is identified by a vanishing
thermal expansion.

\begin{figure}
\includegraphics[width=\linewidth,keepaspectratio]{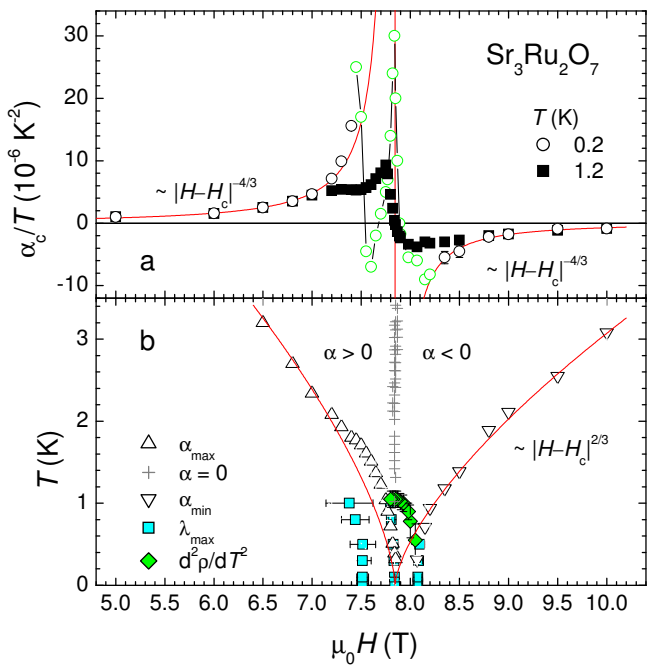}
\caption{\label{fig:double plot}(color online) (a): Field
dependence of the thermal expansion coefficient of
Sr$_3$Ru$_2$O$_7$ as $\alpha_c/T$ vs $\mu_0H$ at 0.2~K (open
circles) and 1.2~K (closed squares). The green open circles
indicate data in the non-Fermi liquid regime,
$\partial^2\alpha_c/\partial T^2 \neq 0$. The red solid lines
show a fit $|H-H_c|^{-4/3}$ with $\mu_0H_c=7.845$~T.
(b): Corresponding ($T,H$) diagram. Triangles and crosses show
positions of thermal expansion extrema and zeros, respectively.
Squares and diamonds indicate positions of peaks in
magnetostriction and in $d^2\rho/dT^2$, respectively, see also
Fig.~\ref{fig:phase anomalies}.}
\end{figure}

Within the theory (\ref{action}) this accumulation of entropy is
expected exactly at $H=H_c$ ensured by the (emergent) reflection
symmetry $\psi \to -\psi$ at $h=0$, leading to the property
$\alpha(-h)= - \alpha(h)$ near the QCEP. The behaviour of the
thermal expansion deriving from (\ref{action}) is summarized in
Fig.~\ref{fig:Theory}. If the system is tuned away from the
metamagnetic transition, $h \neq 0$, the field $\psi$ attains a
non-zero expectation value. At low temperature its response is
non-linear, $\psi \propto |h|^{1/\delta}$, with the mean-field
exponent $\delta = 3$. Thermodynamics is determined by the
fluctuations around this mean-field solution giving rise to
regimes I and II. The cross-over I/II is characterized by
incipient deviation from Fermi-liquid behaviour.  At elevated
temperatures however thermal fluctuations will stabilize the
mean-field potential resulting in a weaker, linear response,
$\psi \propto h$.  In this linear response regime III the
behaviour of $\alpha$ is captured by renormalized mean-field
theory giving a maximum of $|\alpha|$ at the crossover II/III
followed by a steep decline at higher $T$.

\begin{figure}
\includegraphics[width=\linewidth,keepaspectratio]{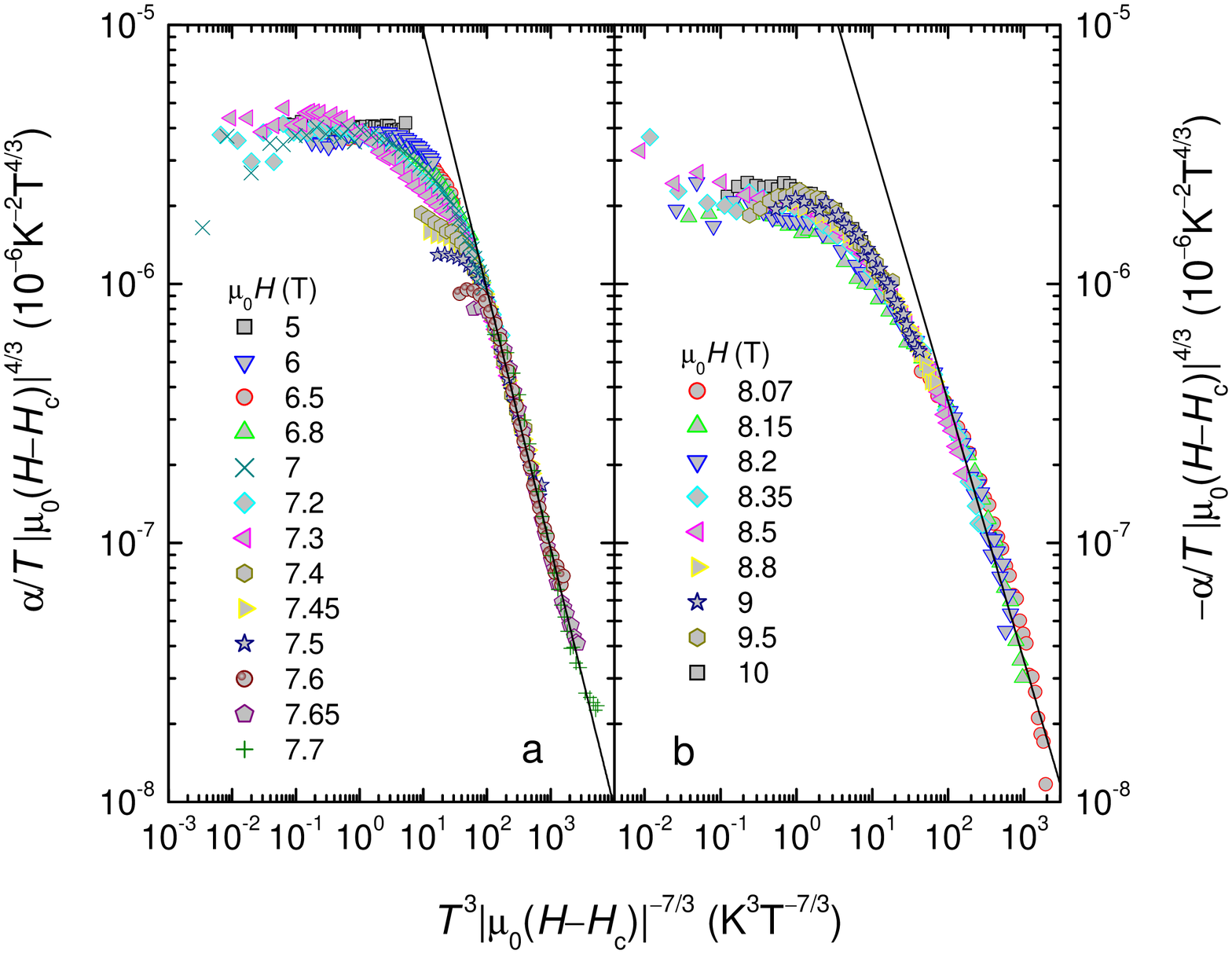}
\caption{\label{fig:scaling} (color online) Linear $c$-axis
thermal expansion coefficient of Sr$_3$Ru$_2$O$_7$ rescaled as
$\alpha_c/T \times |\mu_0 (H-H_c)|^{4/3}$ vs $T^3 \times |\mu_0
(H-H_c)|^{-7/3}$ with $\mu_0 H_c = 7.845$ T (on a double-log
scale) for low-field (a) as well as high-field (b) data. For
$7.4$~T $\leq\mu_0H\leq 8.07$~T data for temperatures below 1.1~K
have been omitted. Lines indicate $\alpha_c \propto |\mu_0
(H-H_c)| T^{-2}$.  }
\end{figure}

For comparison with experiment the magnetic field dependence of
the thermal expansion coefficient, $\alpha_c/T$, is shown in
Fig.~\ref{fig:double plot}a. At elevated temperatures (1.2~K) the
behavior resembles that observed near the metamagnetic cross-over
in CeRu$_2$Si$_2$ \cite{Lacerda89} with a sign change of
$\alpha_c/T$ occuring at $\mu _0 H_c=(7.845\pm 0.005)$~T. Using
this value for the critical field, we can describe the field
dependence inside the Fermi liquid regime by $\alpha_c/T\propto
|H-H_c|^{-\epsilon}$ with $\epsilon=1.35 \pm 0.1$, in agreement
with the expectation for two-dimensional (2D) spin fluctuations
($\epsilon = 4/3$ red solid lines in Fig.~\ref{fig:double plot}a).
Two-dimensional spin fluctuations were also observed in inelastic
neutron scattering \cite{Capogna03} at zero magnetic field which
revealed that they are confined to the RuO$_2$ bilayers of the
crystal structure.
Close to the metamagnetic region, the low-$T$ thermal expansion in
Fig.~\ref{fig:double plot}a shows complicated behavior with three
sign changes in a narrow field interval, which are attributed to
the fine structure near $H_c$ mentioned in the introduction.
Outside this field region, the zeros of $\alpha$ are independent
of $H$, see Fig.~\ref{fig:double plot}b.
The positions of thermal expansion extrema follow a
$|H-H_c|^{2/3}$ dependence as expected in $d=2$ (red line in
Fig.~\ref{fig:double plot}b). However, at lowest temperatures
they do not coincide at a {\it single} (quantum critical end)
point at $T=0$ but merge with the two lines of first-order
transitions. The presence of these first-order transitions might
also be at the origin of the missing symmetry
$\alpha(-h)=-\alpha(h)$ already apparent in Fig.~\ref{fig1}. 

The scaling-plot in Fig.~\ref{fig:scaling} analyzes the full
shape of $\alpha_c$ in more detail; however, omitting data at
temperatures $T\leq 1.1$~K for fields very close to $H_c$. The
thermal expansion has been rescaled such that a data collapse for
the low- and high temperature asymptotes, regime I and III,
should occur for a 2D QCEP. Note that the presence of the
intermediate regime II prevents a full collapse onto a single
scaling curve and deviations in this regime are even
theoretically expected. The scaling behaviour of the raw data is
remarkable given the uncertainties associated with a possible
small linear in $T$ background (see $\mu_0 H = 7.8$~T in
Fig.~\ref{fig1}). Moreover, in the low-field region, $H<H_c$, the
additional anomaly at 7.5 T leads to distortions at low
temperatures. The lines indicate the high-$T$ asymptotes expected
in $d=2$. Whereas in the high-field region the fit is not
compulsory, the behaviour in the low-field region is convincingly
described by two-dimensional metamagnetic fluctuations. Note
however that a recent NMR study \cite{Kitagawa05} indicates that
the full fluctuation spectrum near the critical field might also
contain in addition an antiferromagnetic component.

\begin{figure}
\includegraphics[width=\linewidth,keepaspectratio]{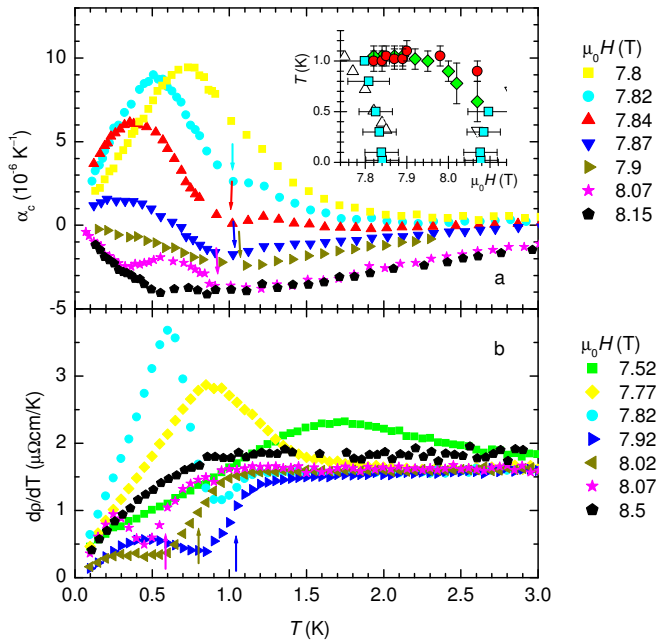}
\caption{\label{fig:phase anomalies}(color online) (a):
Temperature dependence of the thermal expansion coefficient
$\alpha_c$ (a) and resistivity derivative $d\rho/dT$ (b) of
Sr$_3$Ru$_2$O$_7$ at various magnetic fields $H\parallel c$.
Arrows indicate anomalies associated with the entry into the
bounded state. The inset displays the ($T$,$H$) phase diagram
(cf.~Fig.~3b) for the region very close to the metamagnetic
transitions. Red circles, and green diamonds mark positions of
arrows in (a) and (b), respectively.}
\end{figure}

At last, we concentrate on the field regime near 8~T between the
two first-order transitions \cite{Perry04} (squares in inset of
Fig.~\ref{fig:phase anomalies}a) where at low temperatures an
enhanced residual resistivity was observed \cite{Grigera04}. As
shown in Fig.~\ref{fig:phase anomalies} this regime in the
($T$,$H$) diagram is bounded in temperature by a cross-over line
connecting the two first-order transitions where the thermal
expansion and especially the electrical resistivity show
anomalies. Here, the resistivity derivative drops sharply
(cf.~arrows in Fig.~\ref{fig:phase anomalies}); the inflection
points determined from the maxima in $d^2\rho(T)/T^2$ define the
positions of the diamonds in the phase diagram shown in the
inset.
A discontinuous change in thermal expansion, expected when
crossing a phase transition cannot be resolved. By contrast, only
weak minima are observed, whose positions agree well with the
cross-over in the resistivity and signatures in the
dc-magnetization $M(T)$ found previously \cite{Grigera04}.

Because the lines of first-order transitions have slopes pointing
away from the bounded state, it is possible to fine-tune the
field such that this region subsequently is first entered and
then left upon cooling. This leads to distinct anomalies in both
the thermal expansion and the electrical resistivity observed,
e.g., at a field $\mu_0 H = 8.07$~T that are absent at larger
fields when the new state is not entered. The shape of the
bounded region has interesting consequences for its entropic
properties. From the Clausius-Clapeyron relation follows that the
entropy increases when this state is entered upon crossing the
first order transitions at constant temperature. A possible
explanation for the enhanced entropy might be that the soft
fluctuations are not completely cutoff at the crossover
but rather survive within the bounded regime. Note that the
formation of the bounded state at $T\approx1$~K leads only to a
small dip in the thermal expansion (Fig.~\ref{fig:phase
anomalies}a), and a pronounced peak characteristic for the
presence of soft fluctuations also develops within the bounded
region below 1~K at a magnetic field of e.g.~7.84~T.
The observation of an enhanced entropy is also consistent with
the NMR results \cite{Kitagawa05} which showed that quantum
critical fluctuations near the metamagnetic field persist within
the bounded region down to lowest temperatures.

These results place additional constraints on the possible nature
of the bounded region. Its enhanced entropy as well as the
absence of phase transition signatures in the thermal expansion
at the crossovers near 1~K are difficult to reconcile with a
symmetry-breaking mechanism for the phase formation as proposed
in \cite{Grigera04} but it cannot be excluded. If the
incorporation of fluctuation effects in a phase separation
scenario \cite{Honerkamp05} is able to explain the experimental
signatures as e.g.~the shape of the bounded region remains to be
seen. Thus, further theoretical and experimental work is needed
for clarification.


Helpful discussions with S.A. Grigera, A.P. Mackenzie, A. Rosch
and P. W\"olfle are gratefully acknowledged. P.G. thanks the
Deutsche Forschungsgemeinschaft (DFG grant GE 1640/1-1) and Royal
Society (UK) for support of his stay at the St. Andrews
University, M.G. is supported by DFG grant GA 1072/1-1.

\end{document}